# Billion-Scale Similarity Search Using a Hybrid Indexing Approach with Advanced Filtering

*Simeon Emanuilov, Aleksandar Dimov*

Department of Software Technologies, Faculty of Mathematics and Informatics, Sofia University "St. Kliment Ohridski", Sofia, Bulgaria
E-mails: ssemanuilo@fmi.uni-sofia.bg　aldi@fmi.uni-sofia.bg

**Abstract:** *This paper presents a novel approach for similarity search with complex filtering capabilities on billion-scale datasets, optimized for CPU inference. Our method extends the classical IVF-Flat index structure to integrate multi-dimensional filters. The proposed algorithm combines dense embeddings with discrete filtering attributes, enabling fast retrieval in high-dimensional spaces. Designed specifically for CPU-based systems, our disk-based approach offers a cost-effective solution for large-scale similarity search. We demonstrate the effectiveness of our method through a case study, showcasing its potential for various practical uses.*

## 1. Introduction

Similarity search, the task of finding similar vectors, has become a fundamental operation in machine learning, with applications in recommendation engines, semantic search systems, and more [1-3]. As datasets grow to billions of entries, the challenge of performing efficient searches on high-dimensional vectors becomes increasingly complex [4]. This is further compounded by the well-known curse of dimensionality [5], which affects the performance and accuracy of search algorithms as the number of dimensions increases.

Approximate Nearest Neighbor (ANN) algorithms, such as Inverted File Index (IVF) [6] and Hierarchical Navigable Small World (HNSW) [7], have been developed to address scalability and performance issues. IVF segments the search space into smaller areas, called Voronoi cells [8], while HNSW constructs a navigable graph structure for efficient search space traversal. Despite their advancements, these methods often struggle to support complex, multi-dimensional filtering efficiently. This is crucial in practical scenarios where additional criteria beyond vector similarity are required to refine search results [6]. Examples of such scenarios include e-commerce product search and semantic search with filtering and recommendation systems.

　　　　Our key contribution is a disk-based algorithm that integrates dense vectors with discrete filtering attributes, using an enhanced IVF-Flat structure for unified similarity search and filtering. Our approach employs dynamic memory management, loading only necessary index parts into RAM during searches. This design efficiently handles datasets exceeding available memory, scaling to billion-scale data on a CPU server. Combining these elements into a cohesive system provides a practical solution for performing advanced similarity searches on massive datasets without the need for expensive GPU resources for inference.

The paper is structured as follows: Section 2 reviews related work; Section 3 provides an overview of important concepts used throughout this article; Section 4 presents our proposed approach in detail; Section 5 demonstrates the applicability of our method through a case study; and Section 6 concludes the paper with a summary of our findings and potential future directions.

## 2. Related work

Similarity search has witnessed significant advancements in recent years, driven by the increasing prevalence of high-dimensional data in various domains [9]. This section reviews the most relevant literature to our work, focusing on techniques for ANN search and filtering in billion-scale datasets.

2.1. Recent advancements in similarity search algorithms

IVF has been a fundamental approach for ANN search in high-dimensional spaces. Jégou, Douze and Schmid [6] introduced the concept of product quantization, enabling compact representation of high-dimensional vectors and efficient distance computation. This work laid the foundation for many subsequent IVF-based methods.

Baranchuk, Babenko and Malkov [10] further investigated the scalability of inverted indices for billion-scale ANN search, proposing techniques for optimizing the index structure and search procedure.

Johnson, Douze and Jégou [11] demonstrated the effectiveness of GPU-based approaches for billion-scale similarity search, underlining the need for efficient methods to handle massive datasets. While GPU-based approaches have proven to be highly effective, they may not be the most cost-efficient solution for all use cases.

Graph-based methods have emerged as another prominent approach for ANN search. Malkov and Yashunin [7] proposed the HNSW graph, which constructs a multi-layer navigable structure to facilitate efficient nearest-neighbor retrieval. The HNSW method has demonstrated strong performance on various benchmark datasets and has been widely adopted in practice. Later, Yang et al. [12] proposed a hierarchical graph index structure and dual residual encoding scheme to improve the accuracy and efficiency of similarity search on billion-scale datasets.

Recently, Zhang et al. [13] proposed a Hybrid Inverted Index (HI2) that combines embedding clusters and salient terms to accelerate dense retrieval. HI2 aims to improve retrieval effectiveness and efficiency by leveraging both semantic and lexical features.

However, incorporating filtering capabilities into ANN search has received limited attention in the literature. Filtered-DiskANN, proposed by Gollapudi et al. [14], represents a notable effort in this direction, extending the DiskANN system [15] to support simple, one-dimensional filters within a graph-based index. While this work highlights the importance of filtering in practical similarity search scenarios, it is limited in efficiently handling complex, multi-dimensional, SQL-like filter expressions, as stated by the authors [14].

While our work focuses on general similarity search, the proposed hybrid indexing approach could potentially be applied to specific domains like facial analysis. For instance, Al-Dujaili et al. propose a hybrid model for age estimation from facial images using machine learning techniques [16], demonstrating the broad applicability of combining multiple features for improved performance in various tasks.

Several other works have explored various similar aspects, such as real-time updates in ANN indexes [17], distributed indexing techniques for streaming similarity search on billion-scale tweet datasets [18], and industry solutions like pgvector [19] and AnalyticDB [20]. However, these systems often face limitations regarding index size [21], dimensionality, filtering capabilities, or the hardware required [22, 23]. While these advancements have improved similarity search capabilities, scaling these methods to billion-scale datasets presents unique challenges.

2.2. Large-scale similarity search methods

Large-scale similarity search has evolved to meet the challenges of ever-growing datasets. Quantization-based methods, such as Product Quantization, compress high-dimensional vectors to reduce memory requirements and accelerate computations, though they may sacrifice some accuracy. Graph-based approaches like HNSW construct navigable structures for efficient search, offering high accuracy but potentially becoming memory-intensive at a billion scale [24].

Tree-based and Locality-Sensitive Hashing (LSH) methods provide alternative strategies for partitioning the search space. However, they face issues with high-dimensional data [25]. IVF-PQ (Inverted File with Product Quantization) combines inverted file structures with product quantization, balancing memory efficiency and fast search times. FAISS (Facebook AI Similarity Search) [26] offers a comprehensive library implementing many of these techniques, providing efficient similarity search and clustering for dense vectors.

Despite these advancements, efficiently incorporating filtering capabilities into these methods remains a significant challenge for practical applications, often leading to performance degradation or requiring extensive post-processing.

2.3. Vector search algorithms with filtering capabilities

To contextualize our approach, we compare it with several prominent algorithms that attempt to address the filtering challenge. Filtered-DiskANN [14], developed by Microsoft, extends the DiskANN [15] system to support simple, one-dimensional filters within a graph-based index. While efficient, it's limited in handling complex, multi-dimensional filtering expressions.

The popular PostgreSQL extension pgvector [27] implements HNSW and IVF for vector similarity search. Although it performs well for small datasets, it struggles with efficient filtering and index construction for billion-scale collections on typical CPU servers. Similarly, StreamingDiskANN (pgvectorscale [28]) aims to enhance pgvector's capabilities with disk-resident ANN search but faces challenges with large datasets on standard hardware.

To illustrate these limitations, we conducted initial experiments before our large-scale evaluation in Section 5. Using a sample of 15 million 768-dimensional normalized vectors, we attempted to build indexes using both pgvector and pgvectorscale. These attempts were unsuccessful on CPU-based hardware, resulting in system unresponsiveness and excessive processing times [29]. These outcomes underscore the scalability challenges these methods face when dealing with large collections, setting the stage for our proposed approach.

While the aforementioned approaches, methods, and libraries have made significant contributions, there remains a gap in efficient and flexible filtering capabilities for large-scale similarity searches. Our work aims to address this gap by proposing a novel approach that integrates similarity search and multi-dimensional filtering within an optimized IVF-Flat structure.

## 3. Background

Before exploring the details of our proposed approach, we provide a brief overview of the important notations used throughout this paper.

### 3.1. Inverted File index (IVF)

IVF is a fundamental approach for ANN search in high-dimensional spaces [6]. IVF partitions the search space into Voronoi cells (Section 3.3), each associated with a centroid vector. The IVF index consists of two main components:

1. A set of $K$ centroids, denoted as $C = \{c_1, c_2, ..., c_K\}$, where each centroid $c_k \in \mathfrak{R}^D$ represents the center of the $k$-th Voronoi cell, $k$=1, ..., $K$, and $D$ represents the dimensionality of the vectors in the dataset.

2. A set of $K$ inverted lists, denoted as $L = \{L_1, L_2, ..., L_K\}$, where each list $L_k$ contains the identifiers of the vectors assigned to the $k$-th centroid.

During indexing, the identifiers (pointers) of dataset vectors are assigned to their nearest centroids, forming these inverted lists (Section 3.3). At query time, the search is limited to a subset of the most promising inverted lists (Section 4.4).

### 3.2. IVF-Flat

IVF-Flat extends the basic IVF index by incorporating a flat index structure within each inverted list. In our implementation, this flat index is stored on disk, maintaining the actual vectors in a contiguous file layout. The IVF-Flat index inherits the notations from the IVF index, with the addition of the flat index component within each inverted list $L_k$.

### 3.3. Voronoi cells

Voronoi cells are fundamental geometric structures and divide the vector space into regions, each containing all points closer to its associated centroid than to any other centroid. Given a set of centroid vectors $C = \{c_1, c_2, \ldots, c_K\}$, the Voronoi cell associated with a given centroid $c_k$ is defined as

(1) $\qquad V_k = v \in \Re^D \mid d(v, c_k) < d(v, c_j)$ for all $j \neq k$,

where:
- $d(\cdot,\cdot)$ is a distance metric (e.g., Euclidean, cosine similarity);
- $v$ is any point in the $D$-dimensional space;
- $d(v, c_k)$ is the distance between point $v$ and centroid $c_k$;
- $(v, c_j)$ is the distance between point $v$ and any other centroid $c_j$.

3.4. Filtering attributes and conditions

Filters in our approach are defined as additional criteria applied to refine search results based on specific attributes associated with the data points. These attributes encompass a range of metadata types, including but not limited to categorical labels, tags, numerical ranges, and other vector-associated information. For example, in an image search system, filters might include attributes like size, date, or content tags.

In a given dataset of $N$ raw high-dimensional vectors, denoted as $X = x_1, x_2, \ldots, x_N$, where each vector $x_i \in \Re^D$, we use $M$ as the number of filtering attributes, and respectively define a filter vector $a_i = [a_{i_1}, a_{i_2}, \ldots, a_{i_M}]$. These attributes represent the additional metadata that can be used to refine search results.

A set of filtering conditions $F = f_1, f_2, \ldots, f_M$ specifies criteria for one or more of these $M$ attributes. Each component $f_g$, $g = 1, \ldots, M$, of $F$ defines a constraint on the $g$-th attribute across all vectors, utilizing relational operators and values to specify precise filtering criteria. As a practical example, consider a filtering condition that might require that a specific attribute equals a certain value or falls within a particular range.

Each attribute is represented as a fixed-size integer value, facilitating rapid comparisons and bitwise operations. This encoding supports a diverse range of filter types, including exact match queries, range queries implemented via interval trees, and multi-attribute logical operations. To accommodate various data types, we employ one-hot encoding for categorical attributes and adaptive binning techniques for numerical attributes, striking a balance between expressiveness and dimensionality reduction.

During a search operation, the filtering condition $F$ is applied to narrow down the set of candidate vectors, using the filter vector (Section 4.4). The search algorithm ensures that only vectors satisfying all specified conditions in F are considered for the ANN search.

3.5. Vector types

Throughout this work, we use several types of vectors:
- **Core vector** ($x_i$). The original, raw, high-dimensional vector representation of the data point, typically output from a neural network. We use $x_i \in \Re^D$ to denote a core vector, where $D$ is the dimensionality of the embedding space.

- **Attribute vector** ($a_i$). Represents the discrete filtering attributes associated with each data point. We denote the attribute vector as $a_i = [a_{i_1}, a_{i_2}, ..., a_{i_M}]$, where $M$ is the number of filtering attributes.
- **Hybrid vector** ($h_i$). The concatenation of the core vector and the attribute vector. It is represented as $h_i \in \mathcal{R}^{(D+M)}$, combining both the embeddings and the filtering attributes.
- **Query vector** ($q$). Represents the search query, which is a hybrid vector, a result from the concatenation from the core search vector and the search filtering attributes [ $x_{\text{input}}$ || $a_{\text{input}}$] (|| denotes concatenation, see Section 4.4).

It's worth noting that we often use the terms "vector" and "embedding" interchangeably.

## 4. Proposed approach

In this section, we present the proposed approach for a cost-efficient, large-scale similarity search with complex filtering capabilities.

4.1. Constructing hybrid vectors

First, we need to construct the hybrid vector, mentioned in Section 3.5. The hybrid vectors are denoted as $H = \{h_1, h_2, ..., h_N\}$, where each vector $h_i \in \mathcal{R}^{(D+M)}$, $i = 1, ..., N$. Given a dataset of $N$ raw high-dimensional vectors, denoted as $X = \{x_1, x_2, ..., x_N\}$, where each vector $x_i \in \mathcal{R}^D$, and a corresponding set of $M$ filtering attributes, denoted as $A = \{a_1, a_2, ..., a_M\}$, we construct the hybrid vectors as $h_i = [x_i || a_i]$, where $h_i$ is the hybrid vector corresponding to the $i$-th data point, $N$ represents the total number of vectors in the dataset, and $M$ represents the number of filtering attributes available (Fig. 1).

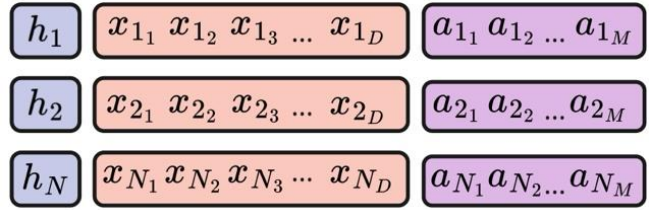

Fig. 1. How the hybrid vectors are constructed

In Fig. 1 each row corresponds to the $i$-th data point and respectively represents:
- $h_1$ to $h_N$ – hybrid vector index.
- $x_{i_1}$ to $x_{i_D}$ – the core embedding (typically coming from a neural network).
- $a_{i_1}$ to $a_{i_M}$ – the filtering attributes.

This unified representation offers several advantages: it eliminates the need for multiple indexing structures, reducing storage and maintenance overhead while providing flexible and dynamic filtering. By combining dense embedding vectors and discrete filtering attributes into a single hybrid vector, we create a compact representation that encapsulates both semantic similarity information and relevant

50

metadata for each data point, allowing for easy modification of filtering conditions without altering the underlying index structure.

4.2. Hybrid index construction

The construction of the hybrid index involves several key steps:

**1. Centroid computation.** K-Means or MiniBatchKMeans [30] clustering is performed on the core vectors $x_i \in \Re^D$ to obtain $K$ cluster centroids, denoted as $C = \{c_1, c_2, \ldots, c_K\}$, where each centroid $c_k \in \Re^D$. These centroids serve as the representatives of the inverted lists and are used to guide the search process.

**2. Vector assignment.** Each core vector $x_i$ is assigned to its nearest centroid $c_k$ based on a distance metric $d(\cdot,\cdot)$, e.g., cosine similarity. This forms the inverted lists, denoted as $L = \{L_1, L_2, \ldots, L_K\}$, where each list $L_k$, $k$ = 1, …, $K$, contains the indices of the core vectors assigned to the $k$-th centroid.

**3. Flat index construction.** For each inverted list $L_k$, the full core vectors are stored. This flat storage approach means that the complete vector data is retained, as opposed to pointers or compressed, quantized representations.

**4. Filter attribute association.** In addition to the core vectors, the index structure maintains the corresponding filter attributes for each vector. These attributes are stored in a manner that preserves their association with the core vectors, allowing for efficient filtering operations during the search process.

The resulting structure (Fig. 2) allows quick identification of relevant clusters during search and enables filtering and precise distance calculations within those clusters.

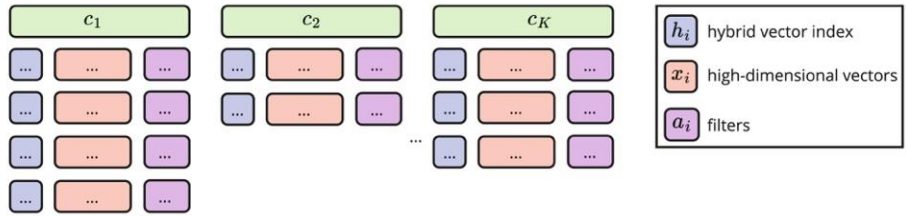

Fig. 2. Centroids, index elements, filters

The number of centroids ($K$) in the IVF-Flat index affects the trade-off between index size, construction time, and search efficiency [31]. A common heuristic is to set $K$ to $N/1000$ for datasets up to 1 million vectors, or $\mathrm{sqrt}(N)$ for larger datasets, where $N$ is the total number of vectors (Section 4.3) [27].

4.3. Centroid selection and search parameters

The number of centroids ($K$) in our hybrid indexing approach and the number of nearest centroids ($T$) selected during the search significantly influence the performance and accuracy of similarity search with filtering.

During index construction, $K$ determines the granularity of space partitioning. A larger $K$ results in finer partitioning, potentially improving search accuracy but increasing index size and construction time. Conversely, a smaller $K$ leads to coarser partitioning, reducing index size but potentially sacrificing some accuracy. We empirically confirm the common heuristics that setting $K$ to approximately sqrt($N$), where $N$ is the total number of vectors, provides a good balance for billion-scale datasets.

In query execution, $T$ affects both accuracy and performance. A small $T$ may miss relevant results, especially when filtering is applied, while a large $T$ can lead to slower search times.

Regarding search complexity, while in the worst case, it can approach $O(N)$, in practice, the combination of centroid-based pruning and efficient filtering typically results in sub-linear search times. We provide empirical results (Section 5) demonstrating the effectiveness of this approach on billion-scale datasets. To address potential memory constraints, a future direction could be a disk-based storage strategy with intelligent caching. Frequently accessed parts of the index are kept in memory, while less frequently used portions are stored on disk.

Also, it's important to note that the optimal $K$ and $T$ can vary depending on factors such as data distribution, dimensionality, and specific filtering requirements. Future work could explore adaptive methods for determining $K$ and $T$ based on dataset characteristics, query patterns, and filter selectivity to further optimize the trade-off between search accuracy and speed.

4.4. Search

Given a query vector $q \in \mathcal{R}^{(D+M)}$, containing an input for the search (core vector), denoted as $x_{\text{input}}$, and list of filtering conditions as an attribute vector $a_{\text{input}}$, our method performs the following steps to retrieve the top-$k$ most similar vectors that satisfy the filtering criteria:

**Step 1.** Construct the hybrid query vector $q_h$ by concatenating the query vector $x_{\text{input}}$ with the representation of the filtering conditions $a_{\text{input}}$, i.e., $q_h = [x_{\text{input}} \| a_{\text{input}}]$.

**Step 2.** Identify the $T$ nearest centroids to the hybrid query vector $q_h$ based on the distance metric on $x_{\text{input}}$ part. This step narrows down the search space to the most promising inverted lists. All centroids should be stored in memory and each of the target centroids is denoted with $c_t$, where $t = 1, 2, \ldots, T$ (Fig. 3).

**Step 3.** Apply the filtering conditions $F$ (as attribute vector) on the $T$ selected inverted lists by using an in-memory structure for the filters, discarding any vectors that do not satisfy the specified constraints in $a_{\text{input}}$. This step ensures that only embeddings meeting the filtering criteria are considered for the next step.

**Step 4.** For each of the filtered results, compute the distances between the input query vector $x_{\text{input}}$ and the vectors in the inverted lists. We leverage optimized BLAS (Basic Linear Algebra Subprograms) routines for efficient matrix operations.

**Step 5.** Merge the filtered results from the $T$ inverted lists and select the top-$k$ most similar vectors based on their distances to the query vector $x_{\text{input}}$.

The choice of $T$, the number of nearest centroids to consider, significantly impacts the trade-off between search accuracy and computational cost. A larger $T$ increases the likelihood of finding relevant vectors but also increases search time and memory usage, while a smaller $T$ offers faster searches at the potential cost of recall.

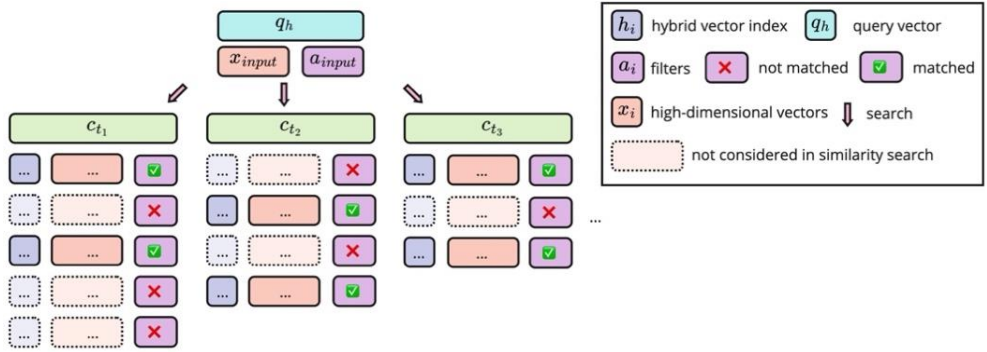

Fig. 3. Searching procedure

One key adaptation in our approach is the dynamic, memory-efficient loading strategy. During a search operation, only the vectors from the $T$ most relevant inverted lists that meet the filtering conditions are loaded into memory. This selective loading significantly reduces memory usage while maintaining high performance.

Our disk-based, dynamically loaded structure effectively manages billion-scale datasets that exceed available RAM. By balancing disk storage with smart, on-demand memory usage, our method achieves an optimal trade-off between scalability, memory efficiency, and search speed, making it well-suited for large-scale similarity search tasks.

4.5. Adding a new vector

In real-world applications, the underlying vector dataset may undergo frequent updates, with new vectors being added. We propose a method for adding a new instance to the index. When a new vector $x_{new}$ and its corresponding filtering attributes $a_{new}$ are added to the dataset, we perform the following steps:

**Step 1.** Construct the hybrid vector $h_{new}$ by concatenating $x_{new}$ with the representation of $a_{new}$, i.e., $h_{new} = [x_{new} \| a_{new}]$.

**Step 2.** Identify the nearest centroid $c_k$ to $h_{new}$ based on the distance metric, calculated from $x_{new}$ part.

**Step 3.** Append $h_{new}$ to the inverted list $L_k$ associated with centroid $c_k$.

**Step 4.** Update the flat index within $L_k$ to include $h_{new}$.

5. Case study

As stated in Section 2.3, our initial experiments demonstrated the limitations of existing methods like pgvector and pgvectorscale when dealing with large-scale datasets. Building upon these initial findings, we conducted a comprehensive case study to demonstrate the effectiveness and efficiency of our proposed algorithm. For this, we utilized the LAION-5B dataset [32], a large-scale multi-modal dataset containing over 5 billion image-text pairs. Specifically, we focused on a subset of 1 billion image embeddings (Laion1B-nolang), each represented as a 768-dimensional vector generated by the CLIP ViT-L/14 model [33], along with their corresponding pre-constructed indexes.

The dataset also includes associated metadata, such as textual captions, URLs, and various attributes like image dimensions and licensing information. We set $T$ (the number of nearest centroids to identify) to 7 as it provides a good balance between search accuracy and computational efficiency.

Table 1 shows the values of parameters used in this case study.

Table 1. Parameters and values

| Parameter | Name | Value |
|---|---|---|
| $N$ | Dataset size | 1 billion ($10^9$) |
| $K$ | Number of centroids (~sqrt($N$)) | 32,000 |
| $D$ | Dimensionality of vectors | 768 |
| $T$ | Number of nearest centroids to identify | 7 |
| $M$ | Number of filtering attributes | 10 |
| $V$ | Average number of vectors per centroid | 31,250 |

We implement the algorithm using Python and the NumPy library for efficient numerical computations. The experiments are conducted on a server with the following specifications: CPU Intel(R) Xeon(R) E-2274G @ 4.00 GHz; 64 GB DDR4 RAM, 1×512 GB NVME HDD + 2×6 TB SATA.

5.1. Hybrid vectors construction

Hybrid vectors were constructed by concatenating 768-dimensional CLIP embeddings [33] with synthetic attribute vectors. For this case study, we append a 10-dimensional vector ($M$) to each CLIP embedding, resulting in hybrid vectors of dimensionality 778.

The attribute vectors are generated to simulate realistic metadata while maintaining a controlled environment. Each dimension of the attribute vector is assigned a random integer value drawn from a uniform distribution in the range [–32768, 32767]. This range is chosen to fully utilize the float16 data type for storage efficiency while maintaining precision.

By using synthetically generated attribute vectors, we can systematically evaluate our algorithm's performance across a wide range of potential metadata configurations.

5.2. Hybrid index construction

Building indexes for billion-scale datasets is computationally intensive. The creators of LAION-5B addressed this challenge [34] using a distributed approach with Autofaiss (wrapper on FAISS [24]), splitting the 9TB embedding collection into 100 parts, and leveraging 10 nodes for parallel processing, with construction time to approximately 16 hours.

For our experiments, we utilized the pre-existing kNN index provided with the LAION-5B dataset and a few processing steps (e.g., merging with filters). To assess scalability, we also tested sci-kit-learn's MiniBatchKMeans [30], constructing an index for our subset in a few hours on a CPU server. The exact time varied with parameters like batch size and iteration count.

While GPU acceleration can significantly reduce indexing time, our focus remains on efficient CPU-based inference. Our proposed algorithm and optimizations enable fast similarity search with complex filtering on billion-scale datasets without requiring GPU resources during the search phase.

5.3. Search

For the search, we used parallel processing by configuring 12 BLAS threads using the OMP_NUM_THREADS environment variable. The sequential execution on a single CPU initially took around 16 s, with the filtering step being the most time-consuming. By utilizing this parallel processing setup, we were able to reduce the search time to approximately 1.428 s.

Table 2. Search performance

| Operation | Time, s |
|---|---|
| Search in centroids | 0.008 |
| Filtering | 1.090 |
| Detailed search in clusters | 0.330 |
| **Total** | 1.428 |

These results highlight the impact of hardware and parallelization [3, 35] on the performance of index creation and search algorithms.

5.4. Discussions and limitations

The case study demonstrates the effectiveness and efficiency of our proposed algorithm. However, certain limitations exist.

Index construction time for billion-scale datasets can be substantial. Potential mitigations include using MiniBatchKMeans [30] for faster clustering or leveraging pre-constructed indexes when available. But the quality of search (i.e., recall) will not be as good as in standard k-Means. Also, some filter attributes may require preprocessing to fit storage constraints (e.g., float32), necessitating normalization or rescaling.

Concurrent searches could also become a bottleneck, as different parts of the index are considered and transferred to memory in the various steps. The proposed approach is more suitable for less frequent access, such as in semantic search or recommendation systems. To overcome this challenge, several solutions can be adopted, including asynchronous request-reply patterns, utilizing a server with more memory or a GPU.

In our experiments, we primarily tested with an exact match for the attribute vector. However, the method is designed to support a range of relational operators allowing for more complex filtering conditions ($F$, Section 3.4). This flexibility enables the system to handle diverse query requirements, though implementing and optimizing for various operators may require additional development and testing.

Despite these challenges, our algorithm provides a practical and cost-efficient solution for similarity search with complex filtering. Future work could address these limitations, exploring adaptive techniques to balance index construction time, search efficiency, intelligent caching, and storage requirements.

## 6. Conclusion

In this paper, we presented a novel algorithm for cost-efficient similarity search with complex filtering capabilities on billion-scale datasets, optimized for CPU inference. Our approach extends the classical IVF-Flat structure by introducing hybrid vectors that integrate dense embeddings and discrete filtering attributes. This method, coupled with a dynamic memory management strategy, enables fast retrieval of relevant vectors while supporting a wide range of filtering conditions.

Our method combines a hybrid vector representation, an efficient IVF-Flat structure, and a similarity search algorithm that utilizes these components to retrieve the most relevant vectors while satisfying complex filtering criteria. Through a case study on the LAION-5B dataset, we demonstrated the practical applicability and efficiency of our approach for large-scale, filterable similarity search.

This work has implications for various applications, including semantic search, recommendation systems, and multimedia retrieval. It also opens avenues for future research, such as adaptive filtering techniques, parallel access improvements, and attribute compression methods. These potential enhancements could further extend our approach to handle an even wider range of real-world scenarios, providing more efficient and effective similarity search capabilities in the era of big data.

*Acknowledgment*: This study is financed by the European Union-NextGenerationEU, through the National Recovery and Resilience Plan of the Republic of Bulgaria, Project No BG-RRP-2.004-0008-C01.